\documentclass[aps,prapplied,reprint,superscriptaddress,longbibliography]{revtex4-2}
\usepackage[utf8]{inputenc}
\usepackage{graphicx}
\usepackage{color}
\usepackage{footnote}
\usepackage{amsmath}

\newcommand{\bb}[1]{{\mathbf{#1}}}
\newcommand{\Iin}[0]{I_{\rm in}}
\newcommand{\Ytil}[0]{\widetilde{\bb{Y}}}
\newcommand{\smc}[0]{\Sigma_{MC}}
\newcommand{\W}[0]{\bb{W}}
\newcommand{\X}[0]{\bb{X}}
\newcommand{\Y}[0]{\bb{Y}}
\newcommand{\MI}[0]{MI_{n,\delta}}
\newcommand{\Par}[0]{P_{n,\delta}}
\newcommand{\rf}[0]{\mathrm}

\newcommand{\OSU}{\affiliation{Department of Physics, The Ohio State University, Columbus, OH 43210, USA}}
\newcommand{\BBN}{\affiliation{Quantum Engineering and Computing, Raytheon BBN Technologies, Cambridge, MA 02138, USA}}
\usepackage[colorlinks=true, linkcolor=blue]{hyperref}

\begin{document}
\title{Reservoir Computing with Superconducting Electronics}
\author{Graham E. Rowlands}\email{graham.rowlands@raytheon.com}\BBN
\author{Minh-Hai Nguyen}\BBN
\author{Guilhem J. Ribeill}\BBN
\author{Andrew P. Wagner}\BBN
\author{Luke C. G. Govia}\BBN
\author{Wendson A. S. Barbosa}\OSU
\author{Daniel J. Gauthier}\OSU
\author{Thomas A. Ohki}\BBN

\date{February 2021}

\begin{abstract}
  The rapidity and low power consumption of superconducting electronics makes them an ideal substrate for physical reservoir computing, which commandeers the computational power inherent to the evolution of a dynamical system for the purposes of performing machine learning tasks. We focus on a subset of superconducting circuits that exhibit soliton-like dynamics in simple transmission line geometries. With numerical simulations we demonstrate the effectiveness of these circuits in performing higher-order parity calculations and channel equalization at rates approaching 100 Gb/s. The availability of a proven superconducting logic scheme considerably simplifies the path to a fully integrated reservoir computing platform and makes superconducting reservoirs an enticing substrate for high rate signal processing applications.
\end{abstract}

\maketitle

\section{Introduction}
Reservoir computing (RC) eschews the conventional paradigms of neural network (NN) construction and training. In a traditional NN the connectivity and connection weights between nodes are carefully designed and individually trained using energy and time-intensive back-propagation methods \cite{Maass2002, Jaeger2004}. In the RC approach, input nodes and computational nodes are assigned fixed and random recurrent connectivity and connection weights, and the output weights are calculated as an explicit function of the reservoir response. There is no need for back-propagation: the reservoir computer is only ever run in inference mode. The same reservoir, which acts as a universal function approximator \cite{Inubushi2017}, can be employed for multiple tasks---even simultaneously---by using different sets of output weights.

Physical reservoir computing sees this recurrent network replaced by a physical system whose dynamical response acts as an effective set of nodes~\cite{Tanaka2019}. The power and convenience of this approach should not be understated. While the operation of ``conventional'' NNs can be hardware accelerated, the hardware platform must often store and recall upwards of $10^8$ individual connection weights. Training requires complex circuitry and is often performed \emph{ex situ} instead using a slow but equivalent software implementation. In a physical reservoir, meanwhile, no memory is required to store internal weight information and there is no need to support back-propagation. Any system can thus be pressed into service as a black-box reservoir so long as it meets certain conditions of nonlinearity and repeatability \cite{Fernando2003}.

\begin{figure}[t!]
  \includegraphics[width=\linewidth]{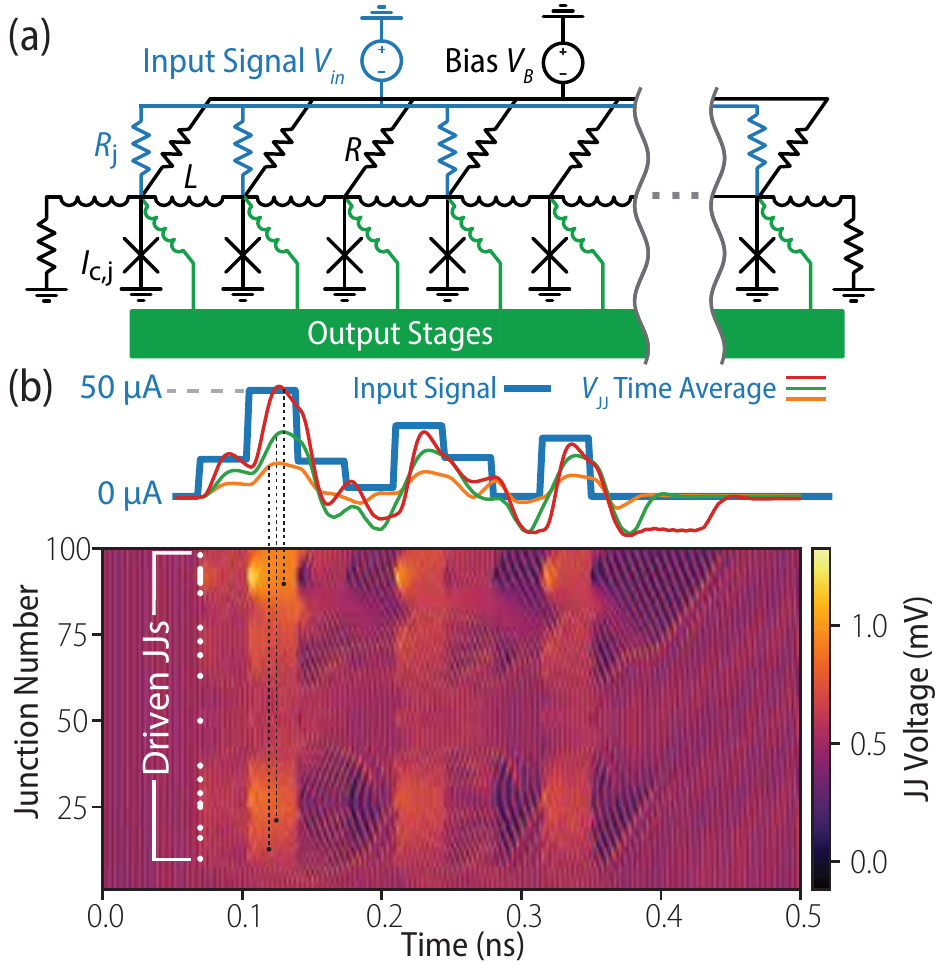}
  \caption{(a) The superconducting reservoir formed by a Josephson transmission line with the input signal distributed to some fraction $f$ of the junctions across individual resistances $R_j$. The junctions are connected to decimating output stages. (b) The reservoir response to the displayed 35-ps sample-and-hold input waveform, showing the voltages across each JJ in a chain of 100. Only twenty percent of JJs (white dots) are driven with the input signal shown above. The time-averaged voltages $V_{jj}(t)$ are shown, for a few indicated junctions, superimposed on the input signal.}
  \label{fig:Device}
\end{figure}

As one might expect, a number of physical RC architectures have been demonstrated:  optical \cite{Duport2012}, optoelectronic \cite{Paquot2012, Larger2017}, spintronic \cite{Torrejon2017}, mechanical \cite{Dion2018}, memristive \cite{Kim2012}, and asynchronous field-programmable gate array (FPGA)-based systems \cite{Canaday2018} among many others \cite{Tanaka2019}. The majority of these hardware platforms use a small number of physical nonlinear elements (often just one), and make clever use of temporal multiplexing and feedback to create virtual nodes that expand the effective dimensionality of the reservoir. Practically speaking, these implementations are complex --- especially those that digitize outputs and subsequently re-synthesize stimuli --- and take a considerable toll on system throughput, which is reduced in direct proportion to the number of virtual nodes.

\section{Reservoir Description}

The design-space of potential superconducting circuit reservoirs is vast, so we focus our attention on a simple Josephson transmission line (JTL) shown in Fig.~\ref{fig:Device}(a) for this study. The JTL is an active transmission line formed by a chain of biased JJs \cite{Likharev1991}. In the typical operation mode, an input pulse from one end of the JTL causes a rapid cascade of junction phase slips that propagate the single flux quantum (SFQ) pulse to the other end.

It has been known for some time that these flux excitations are solitons: a consequence of the JTL's dynamics being governed by a perturbed Sine-Gordon equation in the discrete limit \cite{Fujimaki1987}. Recent work has explored the general RC capacity offered by soliton systems \cite{Marcucci2020, Silva2021}, and invites the question of whether the success of the JTL reservoir is shared with a broader class of systems. After all, the first implementation of a physical reservoir computer was based on complex wave interactions in shallow water \cite{Fernando2003}, one of the canonical systems for the study of soliton dynamics \cite{Korteweg1895, Zabusky1965}.

To operate the JTL as a reservoir, we bias all of the JJs (indexed by $j$) with a global voltage $V_b$ that results in a per-JJ bias currents $I_b = V_b/R$ for bias resistance $R$. For $I_b$ less than their critical currents $I_c$, the JJs actively propagate existing pulses as in typical JTL operation. For $I_b$ greater than $I_c$, the JJs enter oscillatory states whose frequencies are a nonlinear function of the total junction current. Because the JJs modulate each others' currents, a complex dynamical state is achieved. In this regime, the JTL acts, in effect, as an analog liquid state machine \cite{Maass2002}. The JTL is terminated at either end by resistances matched to the JJ shunt resistance, creating an absorbing boundary condition for pulses at the edges of the JTL chain.

We connect the input signal $V_{\rm in}(t)$ to some fraction $f$ of the JJs through additional resistors $R_j$ to produce currents $\Iin^j(t)$. Using a spread in $R_j$ or non-unity $f$ avoids input symmetries that produce spatially homogeneous, and therefore computationally uninteresting, dynamics. Heterogeneity in many other parameters, such as per-junction variations of $I_c$, will also suffice. While natural fabrication-induced spreads in these parameters may be enough to achieve this goal, deliberately engineering asymmetry guarantees success.

\begin{figure}[b!]
  \includegraphics[width=\linewidth]{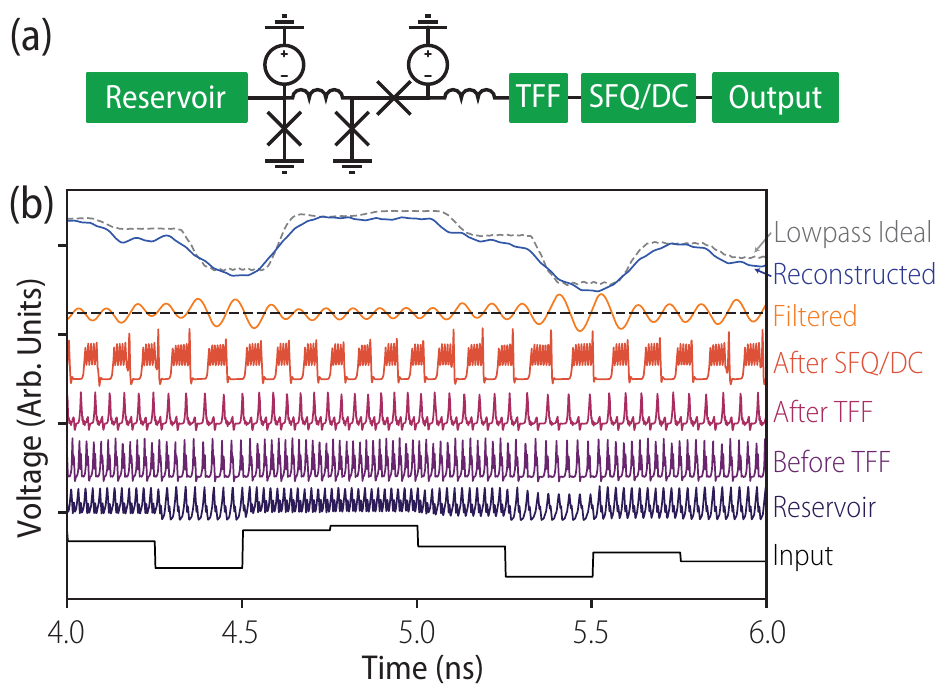}
  \caption{(a) Detailed view of the output stage, which consists of a buffered output junction, a toggle flip-flop, and an SFQ-to-DC converter. (b) The signal as it evolves, from bottom to top, while passing through the output stage in response to the reservoir stimulus shown at the bottom. The threshold for signal reconstruction is shown as the dotted line at the mean of the ``filtered'' response and the resulting reconstruction is compared to the ``ideal'' time-averaged reservoir response.}
  \label{fig:Output}
\end{figure}

The voltage response of an example reservoir with $N=100$ JJs is shown in Fig.~\ref{fig:Device}(b). The locations of driven JJs are indicated, from which wave-like modulations in the $>200$ GHz oscillation frequencies radiate. We take the time-averaged voltage of the JJs, which is related to their average oscillation rates, as the reservoir output quantity. The time-averaged responses for a few junctions are shown in Fig.~\ref{fig:Device}(b) as obtained through a 60-GHz-cutoff low-pass-filter. While proven SFQ circuitry could process these signals on-chip, for a conceptual demonstration it is preferable for input stimuli and data acquisition to be possible with room temperature equipment. The dynamics of this example reservoir (the so-called ``fast'' JTL) are simply too rapid to pass to room temperature through typical transmission lines, so we slow down the dynamics (in a so-called ``slow'' JTL) and implement an SFQ output stage that further reduces the data rate with decimation. 

\begin{table}
  \begin{tabular}{llll}
  \hline \hline
    Architecture & Fast JTL & \multicolumn{2}{c}{Slow JTL} \\ 
    Task & Parity & Parity & Chan. Eq. \\ 
    \hline
    Critical Current $I_c$ ($\mu$A)     & 140 & 50 & 50\\
    Shunt Resistance ($\Omega$)         & 1 & 2 & 2\\
    Total Capacitance $C$ (fF)          & 185 & 670 & 670 \\
    Stewart-McCumber $\beta_c$          & 0.08 & 0.4 & 0.4\\
    Number of JJs $N$                   & 45 & 50 & 50\\
    Coupling Induct. $L$ (pH)           & 10 & 35 & 35 \\
    Bias Current $I_b$                  & 0.8\ $I_c$ & 0.8\ $I_c$ & 1.75\ $I_c$ \\
    Input Resist. $R \pm \Delta R$ ($\Omega$)   & $20\pm 2$ & 20 & 20 \\
    Input Current $\Iin \pm \Delta \Iin$     & $(2.0\pm0.2) I_c$ & $0.29\ I_c$ & $0.32\ I_c$\\
    Input Fraction $f$                  & 1.0 & 0.42 & 0.3 \\
    Data Rate (GS/s)                    & 40 & 5.0 & 5.2\\
    Low-Pass Filt. Freq. (GHz)          & 25 & 13.6 & 8\\
    Sampling Rate (GHz)                 & 400 & 25 & 26.3 \\
    Training Data size                  & 1000 & 1000 & See Note$^*$ \\
    Testing Data size                   & 500  & 1000 & See Note$^*$ \\
    \hline \hline
  \end{tabular}
  \caption{Parameters used for both fast and slow JTL reservoirs for simulations of both parity and channel equalization tasks, unless specifically noted in the main text. The per-junction spreads $\Delta R$ and $\Delta \Iin$ are given as standard deviations. $^*$Training and test data for the channel equalization task are simulated in batches of 5,000 symbols, many of which are run in parallel before being combined to achieve the desired statistics.}
    \label{tbl:circuits}
\end{table}

The circuit parameters for these ``fast'' and ``slow'' variants of the JTL RC, and any differences between these parameters for different reservoir applications, are shown in Table \ref{tbl:circuits}. The output stage for the slow JTL is shown in Fig.~\ref{fig:Output}(a). An output junction and buffer stage condition the reservoir response into discrete SFQ pulses, then a toggle flip-flop (TFF) and SFQ-to-DC converter decimate the pulses by a factor of 4 and output a non-return-to-zero (NRZ) signal whose transitions give a time-encoding of the original signal amplitude and can readily pass through a 12-GHz microwave output chain. The original return-to-zero (RZ) JJ response signals from this NRZ output can be recovered using various methods. We employ a simple procedure of low-pass filtering a chain of impulses located at each zero-crossing. This overall readout scheme is shown in Fig.~\ref{fig:Output}(b), where we compare the final reconstructed output signal to the direct JJ voltages with low-pass filtering (which are the outputs used for the fast JTL).

We model two different schemes of introducing heterogeneity into the inputs schemes: one where all JJs are connected to the input signal with per-JJ bias resistance values $R_j$ drawn from a normal distribution, and another where only a fraction $f$ of the JJs are driven with the input signal through identical $R_j = R$ values. Both schemes result in similar performance, but interestingly the fast JTL performs slightly better in the former case and the slow JTL slightly better in the latter. 

\section{Simulation Results}

We focus on simulating reservoir performance for two workloads. First we perform simulations of both reservoirs (fast and slow JTLs), training them to calculate the parities of sub-sequences in a bit stream. Second, we perform additional simulations for the slow JTL, training it to perform channel equalization on scrambled symbols that have passed through a noisy nonlinear channel with inter-symbol interference. For the case of parity, we find that both reservoirs perform similarly well despite differences in their input and output schemes. In the case of channel equalization we find compelling performance and accuracy, suggesting that these architectures are well suited for complex high-rate signal processing applications.

\subsection{Execution \& Training}

We simulate these circuits using WRspice, an open-source SPICE implementation with sophisticated JJ models \cite{Whiteley2020}.
In both the parity and channel equalization tasks, an input signal is fed through the simulated circuit while its response $\X(t)$ is recorded from $M$ of the junction outputs. The input signal has a sample-and-hold period $\theta$ and a rise/fall time of approximately $0.1\,\theta$. We assume that a total of $Q$ symbols are contained within $\X(t)$, and that the sampling rate of the output yields $K$ samples per interval $\theta$. The memory of the reservoir allows us to solve these problems using only the weights from the current sample-and-hold interval. Accordingly, $\X(t)$ is reshaped to a $(K\,M \times Q)$ matrix whose columns contain the stacked per-symbol responses of all junctions. Using half of the recorded data we perform ridge regression to identify the output weights $\W$ that minimize the quantity
\begin{equation}
    | \Ytil - \W \X |^2 + \alpha | \W |^2,
\end{equation} 
where $\Ytil$ is the truth vector and $\alpha$ a regularization parameter that prevents over-fitting and is chosen to maximize performance. Since we treat each symbol as a separate observation, $\W$ is a simple row-vector of length $K\,M$ that multiplies the columns of $\X$. For channel equalization (with $M=12$ and $K=5$), $\W$ is of length 60, while for parity (with $M=N=45$ and $K=10$), $\W$ is of length 450. With the remaining half of the data, we predict the equalized symbols $\Y = \W \X$ and round them to the nearest levels, \textit{i.e.} $(-1,1)$ in the case of parity and $(-3, -1, 1, 3)$ in the case of channel equalization. Given the small size of $\W$, both prediction and training are computationally straightforward.

\subsection{Parity Benchmark}

The parity task \cite{Dion2018} has become an important benchmark in the RC community as it reveals information regarding the (fading) memory capacity of the system. In this task, given an input sequence of bits $[u_i]$ taken randomly from $\{-1,1\}$, the reservoir computer is used to determine the parity of an $n$ bit sub-sequence that occurred $\delta$ steps in the past. Let $\Par$ be the RC's accuracy in doing so. The \textit{mutual information} is defined as 
\begin{align}
    \MI \equiv& \; \Par\log_2(2\Par)\;+ \\
    &\; (1-\Par)\log_2\left( 2 (1-\Par) \right) \nonumber
\end{align}
while the \textit{memory capacity} $MC_n$ is defined as the sum of $\MI$ over all delay steps $\delta$ for a particular parity order $n$ \cite{Busing2010,Dion2018}. We use $\smc$ --- the summation of all $MC_n$ --- as the overall performance metric for this task. 

First we gauge the performance of the fast JTL, which is capable of processing parity data at a 40 GHz rate. The $[u_i]$ sequence is converted into currents $I_i = \Iin (1+u_i)/2$ (bounded between 0 and scaling factor $\Iin$) that we wish to apply to the reservoir's junctions. Heterogeneity is introduced into this reservoir by randomly assigning the input resistance for the $j$-th junction $R_j$ to a value taken from the normal distribution $\mathcal{N}(R,\Delta R)$ with mean $R=20\;\Omega$ and standard deviation $\Delta R$. The globally applied sample-and-hold voltage waveform $V_{\rf in}(t)$ is constructed from the values $V_i = I_i R$, which results in different per-junction currents for non-zero $\Delta R$.

For training, $V_{\rf in}(t)$ is applied to the reservoir to produce the output signal $\X(t)$. Using ridge regression, we identify the per-order weights $\W_n$ such that the predicted parities $\Y_n = \W_n \X$ best reproduce the desired response $\Ytil_n$. A set of parameters that produces good performance is shown in Table~\ref{tbl:circuits}. In particular, we find that the system ``prefers'' a sub-threshold bias current $I_b=0.8\ I_c$ and input current scaling $\Iin > 0.2\ I_c$ that results in the reservoir repeatedly entering and leaving voltage stage oscillations. The performance improves monotonically as $\Iin$ increases further, subject to current-carrying limitations of the circuit. 

\begin{figure}[t!]
\includegraphics[width=0.9\linewidth]{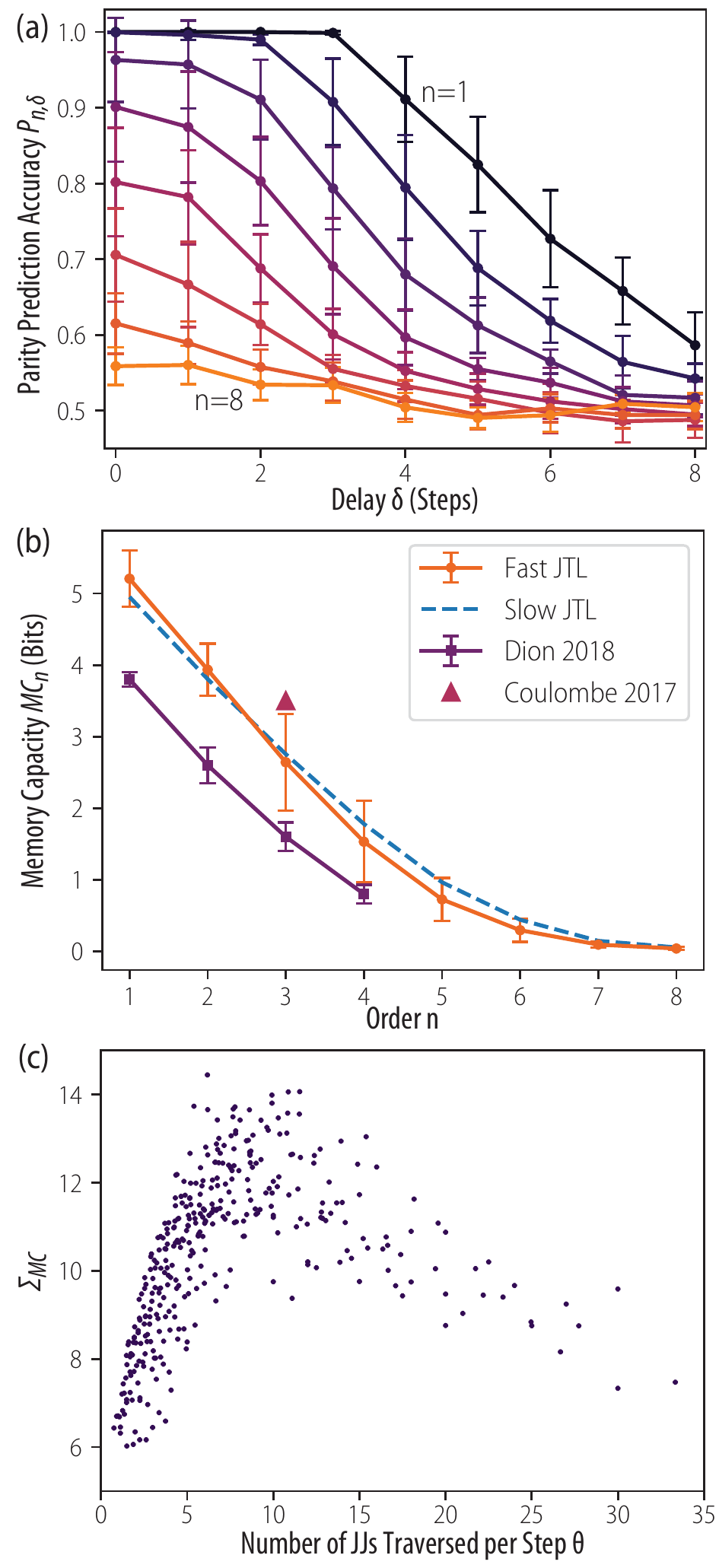}
  \caption{(a) Test accuracy $\Par$ as a function of delay $\delta$ for the fast JTL reservoir described in the text. Parity orders of each trace start from $n=1$ at the top and end with $n=8$ at the bottom. (b) Memory capacity of the fast and slow JTL chains compared to results from single \cite{Dion2018} and coupled \cite{Coulombe2017} mechanical oscillators. For clarity error bars (which show one standard deviation) are suppressed for the slow JTL results, but they are of a similar magnitude to those of the fast JTL data. (c) $\smc$ versus the number of JJs traversed per time step $\theta$, showing a peak at 9 JJs/$\theta$. The data are found for a wide variety of $\theta$, $I_b$, and $L$.}
  \label{fig:Parity}
\end{figure}

We investigate the dependence of $\smc$ on the number of degrees-of-freedom (DOF) in the reservoir by varying both $N$ and $\Delta R$, finding better performance for more DOF: \textit{i.e.}, for larger $N$ and $\Delta R>0$. As noted previously, all JJs in the JTL act synchronously when there is no resistance variation ($\Delta R=0$). All results presented here are for $N=45$ and $\Delta R/R = 0.1$, which amounts to a small and easily fabricated circuit.

Figure \ref{fig:Parity}(a) shows a typical test accuracy $\Par$ of the reservoir as function of $\delta$. The error bars are determined from 10 repeated simulations in which the $R_j$ values are randomly reassigned. As expected, $\Par$ decreases with increasing $\delta$ and $n$, and becomes no better than a random guess as either $\delta$ or $n$ approaches 8. We note that one can construct alternate reservoir geometries that perform well for $\delta \gg 8$ \cite{Barbosa2021}, though memory capacity is poorly defined for such systems.

The parities are converted into memory capacity $MC_n$ and shown in Figure~\ref{fig:Parity}(b), and compared to results from mechanical oscillator reservoirs. The memory capacity of the JTL chain is comparable to that of 400 coupled mechanical oscillators \cite{Coulombe2017} (which is nearly analogous to our system in terms of the governing equations) and that of a single mechanical oscillator having 300 virtual nodes \cite{Dion2018}, while operating at around a million times higher data rate.

We seek a better understanding of how the various circuit parameters in Table~\ref{tbl:circuits} contribute to the effectiveness of the reservoir. It is known, for example, that achieving the desired ``fading memory'' property of a reservoir requires tuning the relative timescales of input data and internal reservoir dynamics \cite{Appeltant2011}. In the context of a time-multiplexed reservoir computer with a single physical oscillator, the sample-and-hold time $\theta$ directly controls these relative timescales. This is often framed as $\theta$ controlling the degree of coupling between virtual nodes, where some optimal value gives a balance between over- and under-coupling. In the JTL reservoir computer, the situation is more complex, because the dynamic interactions between many oscillators provide the RC faculty rather than the dynamics of a single oscillator (as evidenced by the ineffectiveness of homogeneous excitation). We might therefore expect the interaction timescale is what must be matched to the input data timescale.

In Figure~\ref{fig:Parity}(c) we show a parametric plot of $\smc$ versus the number of JJs traversed by pulses during $\theta$, which gives a rough estimate for the interaction timescale between JJs. The data points are taken from simulations over a wide span of $\theta$, $I_b$, and $L$, where higher $L$ and lower $I_b$ result in slower pulse propagation. Accordingly, we capture a wide variety of relative timescales, finding that $\smc$ reaches a peak at the propagation speed of around 9--10 JJs per $\theta$. This suggests that $\theta$ must indeed be matched to the interaction timescale. Further study is needed to develop a more accurate metric for the interaction time between oscillators; nevertheless the propagation time sets a useful lower bound because it ignores the actual response time of the oscillators.

Finally, we gauge the performance of the slow JTL on the parity task. Rather than using a spread in $R_j$ as the source of reservoir heterogeneity, the slow JTL assumes that only some fraction $f$ of the JJs are connected to the input signal $V_{\rf in}(t)$. The reservoir output $\X(t)$ is taken from the decimating chains shown in Fig.~\ref{fig:Output}(a). As with the fast JTL, we find better performance with sub-threshold bias $I_b = 0.8\,I_c$ and $\Iin = 0.29\,I_c$ that causes the reservoir to switch between oscillatory and non-oscillatory regimes. The remaining circuit parameters are shown in Table~\ref{tbl:circuits}. The slow JTL's $MC_n$ values, averaged over 10 realizations of the connectivity, are shown in Fig. \ref{fig:Parity}(b) alongside the fast JTL results. We see nearly identical performance despite notable differences in the input and output strategies for the two reservoir variants, highlighting the robustness of the superconducting platform and of reservoir computing itself. Simulations of the slow JTL produce results that are measurable at room temperature at the expense of reduced throughput: 5 GS/s instead of 40 GS/s but still fast relative to other reservoir implementations, as we discuss in Sect.~\ref{sec:discussion}.

\subsection{Channel Equalization}

\begin{figure}
  \includegraphics[width=1.0\linewidth]{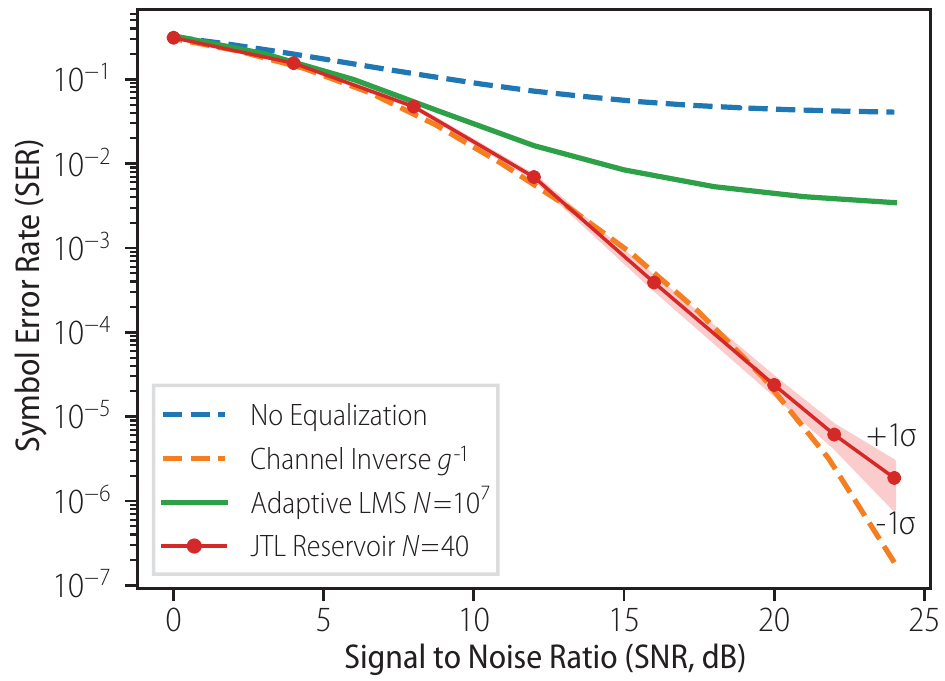}
  \caption{The reservoir's SER vs. the SNR of the channel, with standard deviations shown in the shaded region. The SNR is defined as the ratio of the signal power between two adjacent amplitudes (i.e. -1 and 1) to the power of the AWGN. The performance of an ALMS equalizer trained on $10^7$ points is shown for comparison, as are the results from applying the inverse channel transformation and applying no equalization at all. }
  \label{fig:SER}
\end{figure}

Channel equalization is the process of recovering the symbols that are degraded during transmission over a particular channel, and is of immense technological importance for communications systems. We model a noisy channel exhibiting inter-symbol interference and nonlinear receiver saturation effects that is commonly used in the reservoir computing literature \cite{Jaeger2004}, and assume a four-level pulse amplitude modulation scheme (PAM-4). 

Random PAM-4 sequences $[v_i]$ are generated from the values $(-3, -1, 1, 3)$ and converted into signals $v(t)$ with sample-and-hold time $\theta$ and a 10 ps rise/fall time. Next, $v(t)$ is subjected to the channel transformation function $g$ and additive Gaussian white noise (AWGN) to produce $v'(t) = g(v(t)) + \mathcal{N}(0,\sigma)$ with standard deviation $\sigma$ that is determined by the signal-to-noise ratio (SNR) of the channel. As with the parity task, the input voltage $V_{\rf in}(t)$ is scaled so that it produces currents across the junctions within the interval $[0, \Iin]$. 

Motivated by future measurements of fabricated reservoirs, we run simulations using the slow JTL reservoir described above. Optionally, we can record only the outputs from $M$ (evenly-spaced) junctions rather than then entire chain. Using the same procedure as before, we find the weights $\W$ such that $\Y = \W \X$ best reproduces the optimal response $\Ytil \equiv [v_i]$. The equalization performance is expressed in terms of the symbol error rate (SER) for any elements of $\Y$ that do not match the elements of $\Ytil$. 

We find the reservoir performs well for input fractions $f=$10--40\%, and that its performance does not degrade when as few as every fourth JJ ($M=12$) is included in $\X$. A smaller or larger number of JJs can be used in the reservoir, though the performance begins to degrade rapidly with fewer than 20 JJs. We perform a crude optimization of the reservoir's operating conditions with successive 1D parameter sweeps, finding that bias current $I_b = 1.75\,I_c$, input signal amplitude $I_{\rf in} = 0.32\,I_c$, and sample-and-hold time $\theta=190$~ps produce good performance. More exhaustive hyperparameter searches can be performed with techniques such as Bayesian optimization~\cite{Griffith2019}, but we emphasize that the reservoir performance is robust and is a slowly varying function of all the parameters mentioned above.

The equalization performance is summarized in Fig.~\ref{fig:SER} for SNRs from 0--24~dB, with SERs reaching $10^{-6}$ at the end of this range. The sequence length is adjusted to ensure that the uncertainty in the SER, as calculated from the standard deviation of the corresponding $\beta$-distribution, remains small. We compare to an adaptive least mean squares (ALMS) equalizer trained on $10^7$ symbols, which performs poorly given the exaggerated non-linearity of the channel. We also compare to equalization performed directly with the channel inverse transformation function $g^{-1}(v'(t))$, which nevertheless cannot correct for AWGN. Remarkably, we find that the reservoir performs near this level, suggesting that it has learned to implement $g^{-1}$. In principle the reservoir can actually perform better than $g^{-1}$ since it may learn details of the modulation scheme and noise spectrum during training that can be used to construct a more noise-tolerant equalization strategy. Also remarkable is the data rate: 10.5 Gb/s (5.2 GS/s for 2-bit symbols). Because these reservoirs have been intentionally slowed to enable off-chip post-processing, PAM-4 equalization rates of 80-100 Gb/s should be possible using the fast JTL implementation. 

We highlight an intriguing difference between the parity and channel equalization reservoir parameters: the former is more accurate when the input signal $I_b + I(t)$ repeatedly crosses $I_c$, while the latter is more accurate for inputs that always remain above $I_c$. This suggests that crossing $I_c$ affords an advantage in the overall signal range (given the large change in the time-average voltage across that boundary) that is advantageous for processing the binary-valued inputs for the parity tasks, but overwhelms the continuum of smaller features found in the channel equalization inputs.

\section{Discussion \& Conclusions\label{sec:discussion}}

The ability of the slow JTL reservoir to perform admirably in both parity and channel equalization tasks reaffirms one of the central tenets of RC: a system may be tailored to new tasks merely by finding the appropriate new output weights $\W$. Achieving optimal performance for different tasks does require adjusting $I_b$, $I_s$, and the output filtering, as seen in Table~\ref{tbl:circuits}. As mentioned above, the reservoir appears to favor inputs which cross $I_c$ when a binary input encoding is used (as for parity), while this strategy appears less effective for a continuous input encoding (as for channel equalization). These two regimes will also result in notably different power consumption: for randomly generated inputs bracketing $I_c$ the reservoir dissipates approximately half the power as in the entirely above-threshold regime.
 
The performance of the JTL reservoirs is impressive: producing raw outputs at 5 GS/s and 40 GS/s for the slow and fast JTLs, respectively. These figures are not of particular merit, however, without the ability to perform inference at these same rates. The majority of physical reservoir computer realizations presently leave this final step for offline post-processing, severely bottle-necking their performance. The engineering challenges of achieving weight multiplication at native reservoir rates should not be understated. JTL reservoirs are, fortunately, constructed from components that usually comprise single flux quantum (SFQ) superconducting computing circuits operating with clock speeds approaching 100 GHz \cite{Likharev1991}. One could exploit this compatibility to simply count output pulses within a certain interval and perform weight multiplication directly, rather than having to engineer a sophisticated digitizing architecture. Multiplication in RSFQ has so far been demonstrated (in a single multiplier) at 48 billion operations per second (GOPS) for 8-bit signed integers~\cite{Nagaoka2019}, and at 20 GOPS for 16-bit floating point numbers~\cite{Peng2015}. We have verified that an 8-bit reservoir output encoding retains most of the JTLs' performance, suggesting that integrated RSFQ circuits could readily perform inference at the full rates of both the fast and slow JTL architectures. 

To contextualize this potential performance, we compare to other RC architectures. The fastest reported prediction rates (to out knowledge) are from reservoir computers based on autonomous Boolean logic, which furnish predictions at $180$ MS/s~\cite{Haynes2015, Canaday2018, Rosin2015}. Most other systems presently rely on offline prediction using stored raw outputs from the reservoirs. Optical and optoelectronic reservoir computers with delay-lines typically produce raw outputs at $<100$~MS/s owing to the overhead from large numbers of virtual nodes. A notable exception demonstrates raw outputs at 1.1 GS/s~\cite{Brunner2013}. Non-delay optical systems such as waveguide-based reservoir computers~\cite{Vandoorne2014} have shown some of the highest speeds: 12.5 GS/s for raw outputs. Performing inference at the natural rates of these reservoirs would be difficult given the need for optical-to-electrical conversion, digitization, and finally weight multiplication. By contrast, the JTL's outputs can be directly digitized with simple pulse-counting circuits and piped in parallel to the fast multipliers discussed above.

On a more fundamental level, this work serves as a validation of the predicted RC capability inherent to soliton-supporting physical systems \cite{Marcucci2020, Silva2021}, and is expected to engender further research into the origins of success for reservoir computing on different physical substrates.

\bibliography{the_whole_lib}

\end{document}